\documentclass[twocolumn]{aastex631}

\def\tbin{t_{\rm bin}}
\def\abin{a_{\rm bin}}

\newcommand{\MSun}{{\rm M}_{\sun}}

\renewcommand{\tbin}{t_{\rm bin}}

\usepackage[dvipsnames]{xcolor}
\usepackage{amsmath}
\usepackage{CJK}
\usepackage{tabularx}
\usepackage{array}

\begin{document}
\begin{CJK*}{UTF8}{gbsn}

\title{MRI Turbulence versus $\alpha$-Viscosity in Circumbinary Disks: Implications for Gravitational-Wave Dephasing and Accretion Variability}

\correspondingauthor{Mudit Garg}
\email{mg9113@nyu.edu}

\author[0000-0002-9032-9103]{Mudit Garg}
\affiliation{Center for Cosmology and Particle Physics, Physics Department, New York University, New York, NY 10003, USA}

\author[0000-0002-2624-3399]{Yan-Fei Jiang}
\affiliation{Center for Computational Astrophysics, Flatiron Institute, 162 Fifth Avenue, New York, NY 10010, USA}

\begin{abstract} 
Our understanding of gas torques on massive black hole binaries (MBHBs) relies on hydrodynamical (HD) simulations of circumbinary disks (CBDs) in which an $\alpha$-viscosity models angular momentum transport. However, CBDs around MBHBs near merger are hot, magnetized plasmas where transport is driven by magnetohydrodynamical (MHD) turbulence sustained by the magnetorotational instability (MRI). We test how well an $\alpha$-viscosity captures MRI turbulence with 3D global \texttt{Athena++} simulations of a locally isothermal CBD around an equal-mass, circular MBHB, run in HD and MHD from the same initial torus on identical grids, so that only the transport mechanism differs. Excising the region within twice the binary separation, $r<2~a_{\rm bin}$, allows $1000$ orbits of evolution, several times longer than previous MHD simulations. Both disks deliver the same accretion rate, and the gas beyond $2~a_{\rm bin}$ exerts a positive torque in both, yet the MHD torque is $3.5$ times larger and $3$ times more variable. Near the midplane, transport is similar in the disk body but three times weaker in MHD at the cavity edge, where $2$-$3$ times more mass piles up in a more eccentric cavity, raising the torque. The accretion rate is modulated mainly at twice the binary frequency in HD, but by a persistent lump at the cavity edge and its beat with the binary in MHD. The outer-disk torque therefore implies a contribution to the gas-induced phase shift in the gravitational-wave signal of LISA binaries $3.5$ times larger than in $\alpha$-disk models, and a different variability signature for LSST and Roman.

\end{abstract}
\section{Introduction}\label{S:Intro}

The adoption of the Laser Interferometer Space Antenna (LISA; \citealt{Colpi2024}) mission, together with the in-development TianQin \citep{Li2025} and Taiji \citep{Gong2021} missions, opens a new window for gravitational-wave (GW) astronomy in the millihz band, probing $10^4$-$10^7~\MSun$ systems that are inaccessible to ground-based GW observatories. Among LISA's primary sources are near-equal-mass massive black hole binaries (MBHBs), which are expected to form as a natural byproduct of galaxy mergers \citep{Begelman1980}, since most galaxies host a central MBH \citep{Kormendy2013}. MBHBs can spend up to several years of their inspiral in the LISA band before coalescence, producing thousands of observable GW cycles and signal-to-noise ratios (SNRs) of up to $\mathcal{O}(10^3)$, which potentially allow detection out to redshift $z\sim15$. These observations will therefore enable LISA to probe new regimes of astrophysics \citep{AstroWG2023}, cosmology \citep{CosmoWG2023}, and fundamental physics \citep{FunWG2022}.

Despite this promise, an open question is how MBHBs shrink to the $\sim$milli-pc separations at which GW emission can drive them to merger within a Hubble time. Following a galaxy merger, dynamical friction \citep{Chandrasekhar1943} can efficiently bring the two MBHs down to pc-scale separations at the center of the remnant galaxy. Below this scale, however, dynamical friction becomes inefficient once the binary is bound and the stellar mass enclosed within its orbit becomes comparable to the binary mass, while GW emission becomes efficient only at much smaller separations, leading to the so-called ``final parsec problem'' \citep{Milosavljevic2003}. Over the past two decades, significant theoretical progress has been made on this problem, mainly through improved modeling of the stellar potential \citep[see, e.g.,][]{Preto2009,Khan2011,Vasiliev2015} and of gaseous accretion disks \citep[see, e.g.,][]{MacFadyen2008,Mayer2013,Lai2023,Cattorini2024} in galactic nuclei. Given their mass range, LISA MBHBs are expected to reside predominantly in gas-rich, low-mass galaxies, which provide a natural environment for the formation of a circumbinary disk (CBD) around the binary \citep{AstroWG2023}. Moreover, the interaction of gas with MBHBs can produce electromagnetic (EM) counterparts across the spectrum \citep{Bogdanovic2022,DOrazio2023}, which may be detectable by facilities such as the ongoing Vera C. Rubin Observatory Legacy Survey of Space and Time (LSST; \citealt{LSST2019}) and the recently launched Nancy Grace Roman Space Telescope \citep{Haiman2023}. Finally, even well within the LISA band, the gas disk can remain coupled to the binary until only a few days before merger \citep{Dittmann2023,Garg2026b}, potentially imprinting an observable dephasing on the GW signal \citep{Garg2022,Garg2024b}~-- if left unmodeled, this dephasing can bias tests of general relativity (GR; \citealt{Garg2024d}).

In the literature, two main approaches have been used to study CBDs. The first is hydrodynamical (HD) simulations \citep{MacFadyen2008,Munoz2019,Tiede2020,DOrazio2021,Zrake2021,Franchini2022,Dittmann2022,Siwek2023,Duffell2024}, in which angular momentum transport by turbulence is not modeled self-consistently but is instead parametrized by an effective $\alpha$-viscosity \citep{ShakuraSunyaev1973}, the gas follows a locally isothermal equation of state such that the disk aspect ratio $h_0 \equiv H/r$ ($H$ being the disk scale height and $r$ its orbital radius) is fixed, and gravity is Newtonian. Furthermore, these simulations are typically 2D and assume a fixed binary orbit. As a result, they are computationally inexpensive, can be evolved for hundreds to thousands of binary orbits, and allow for broad surveys of the disk and binary parameter space. These studies have mapped, as a function of the binary mass ratio, orbital eccentricity, and $h_0$, whether a CBD shrinks or expands the binary orbit and whether it damps or excites its eccentricity, providing a first-order understanding of disk-driven binary evolution and of the EM signatures expected before GW emission takes over. Only a few studies have followed the GW-driven inspiral with 3D simulations in which the binary orbit evolves self-consistently \citep{Franchini2024,Garg2025,Garg2026b}; these confirm some of the predictions of 2D fixed-orbit HD simulations but reveal quantitative and qualitative differences in others.

The second approach is 3D magnetohydrodynamic (MHD) simulations that self-consistently include magnetic fields, which are expected to be dynamically important because the gas flows around compact MBHBs are hot, magnetized plasmas. As a result, angular momentum transport arises self-consistently from turbulence driven by the magnetorotational instability (MRI; \citealt{Balbus1991}), which is thought to be the primary mechanism of angular momentum transport in accretion disks, rather than from a parametrized viscosity. Most of these simulations incorporate full general-relativistic dynamics \citep{Noble2012,Farris2012,Gold2014,Noble2021,Combi2022,Ruiz2023,Avara2024,Ennoggi2026} to study the final stages of the binary inspiral, usually the last few tens of orbits before merger, although Newtonian MHD simulations have also been used to follow the binary at wider separations \citep{Shi2012,Shi2015}. However, these simulations are computationally expensive and are therefore generally limited to a few hundred binary orbits at most \citep[e.g.,][]{Noble2021}, shorter than the viscous timescale of the CBD, such that the disk may not even reach a quasi-steady state, in which the accretion rate and the torque on the binary still evolve slowly but their ratio has settled (Section~\ref{S:Post}). For the thermodynamics, they typically employ an ad hoc cooling function that keeps the disk at a prescribed thickness (i.e., constant $h_0$), analogous to the locally isothermal assumption in HD simulations. Modeling the thermodynamics self-consistently requires radiation MHD (RMHD) simulations, in which the disk sets its own temperature through the balance of heating and radiative cooling, and the emergent radiation, which is what EM facilities will actually observe, is computed directly. Such simulations are well established for accretion onto single black holes \citep{Jiang2019a,Jiang2019b,Curd2023}, but have only recently been performed for gas flows around MBHBs \citep{Tiwari2025b,Tiwari2025a,Chan2025}.

In this work, we perform a controlled comparison between 3D HD and MHD simulations of a CBD around an equal-mass binary on a fixed circular orbit, differing only in the mechanism of angular momentum transport: a parametrized $\alpha$-viscosity in HD versus MRI-driven turbulence in MHD. Both sets of simulations are evolved with the same code (\texttt{Athena++}; \citealt{Stone2020}) on identical grids, starting from the same initial torus and adopting the same locally isothermal thermodynamics. By excising the binary from the computational domain, we evolve each simulation for $1000$ binary orbits, to our knowledge several times longer than any previous MHD simulation of a CBD, which span $\lesssim400$ orbits \citep[e.g.,][]{Shi2012,Noble2012,Shi2015,Noble2021,Tiwari2025a,Kim2026}. Our goal is to determine which predictions of HD simulations remain robust when angular momentum transport is instead driven by MRI turbulence. The runs are Newtonian, at fixed separation and with $h_0=0.1$, so they describe binaries wide enough that the viscous time at the cavity edge is shorter than the GW inspiral time and relativistic precession is negligible.

\section{Numerical Setup}\label{S:NumSetup}
We consider an equal-mass MBHB with total mass $M_{\rm bin}$ and separation $\abin$ on a fixed circular orbit in the plane $\theta=\pi/2$, with $G=M_{\rm bin}=\abin=1$, so $\Omega_{\rm bin}=1$ and $\tbin=2\pi$, embedded in a locally isothermal, prograde CBD with aspect ratio $h_0=0.1$. We excise the region $r<2$ around the binary and model each MBH as a point mass $M_{\rm bin}/2$ moving on the circular orbit, as in \citet{Tiwari2025a}. The boundary sits at $2\abin$, rather than at $0.8\abin$ as in \citet{Shi2012} and \citet{Tiwari2025a}, because the time step scales as $r^{3/2}$ on the logarithmic grid; this is what makes affordable the $1000$-orbit integrations that the torque needs to reach a quasi-steady state (Section~\ref{S:Result}), the resolution study (Appendix~\ref{A:L2C}) and the boundary variants (Appendix~\ref{A:boundary}). The price is that the minidisks, the accretion torque and the $m=2$ resonance are excluded (Section~\ref{S:Discussion}). We write $\mathbf r_{{\rm BH},i}(t)$ for the position of MBH $i$ and $\mathbf g_i$ for its gravitational acceleration.

We solve the equations of isothermal hydrodynamics (HD run) and ideal MHD (MHD run) with \texttt{Athena++} \citep{Stone2020}, using its standard second-order finite-volume scheme with constrained transport to keep $\nabla\cdot\mathbf B=0$; the two runs share this scheme and differ in the Riemann solver (Roe for HD, HLLD for MHD), in the explicit viscosity of the HD run and in the magnetic field of the MHD run. The HD run adds an explicit \citet{ShakuraSunyaev1973} viscosity, $\nu(r)=\alpha_{\rm vis}~c_s(r)~h_0r$ with $\alpha_{\rm vis}=0.05$. The isothermal solver uses a single sound speed $c_{s,\rm iso}=c_s(2)\approx0.0707$; we recover the locally isothermal profile $c_s(r)=h_0\sqrt{GM_{\rm bin}/r}$, i.e.\ $P_{\rm eff}=c_s^2(r)\rho$, by adding the acceleration
\begin{align}\label{eq:giso}
  g_{{\rm iso},r} &= -\frac{1}{\rho}\frac{\partial}{\partial r}\Big[\big(c_s^2-c_{s,\rm iso}^2\big)\rho\Big],\\
  g_{{\rm iso},\theta} &= -\frac{c_s^2-c_{s,\rm iso}^2}{r\rho}\frac{\partial \rho}{\partial \theta},\nonumber\\g_{{\rm iso},\phi}&=0\nonumber,
\end{align}
so that $-\nabla P+\rho\mathbf g_{\rm iso}=-\nabla P_{\rm eff}$ in $r$ and $\theta$. We set $g_{{\rm iso},\phi}=0$ to avoid a $\sin\theta\to0$ polar instability; the azimuthal pressure gradient therefore retains $c_{s,\rm iso}^2$, which exceeds $c_s^2(r)$ by a factor $r/2$ --- identical in both runs, and affecting only non-axisymmetric structure, for which it acts as an effective aspect ratio $h_0\sqrt{r/2}\approx0.17$ at $r=6$. Mass sources are the density floor $\rho_{\rm floor}=10^{-14}$ (both runs) and, in the MHD run, an Alfv\'en-speed ceiling $v_{A,\rm ceil}=1$ that prevents prohibitively small time steps in the polar funnel: where $|\mathbf B|/\sqrt\rho>v_{A,\rm ceil}$ we raise $\rho$ to $|\mathbf B|^2/v_{A,\rm ceil}^2$ at fixed velocity.

The initial state is a constant-angular-momentum, isothermal Papaloizou--Pringle torus \citep{Papaloizou1984,Hawley2000} with $l_0=\sqrt{GM_{\rm bin}r_{\rm torus}}$, pressure maximum at $r_{\rm torus}=6$, peak density $\rho_0=10^{-8}$ and sound speed $c_s(r_{\rm torus})$, embedded in a static atmosphere at $\rho_{\rm floor}$. It is an exact equilibrium only for a globally isothermal gas in a point-mass potential, so under Eq.~\eqref{eq:giso} and the binary potential it relaxes early in both runs. A reproducible (seed 1) 1\% density perturbation seeds instabilities. The MHD seed field is a single poloidal loop following the density contours, $A_\phi\propto\max(\rho/\rho_0-0.2,0)$, as in \citet{Tiwari2025a}, who use a $0.5\rho_0$ threshold and set the field so that the average plasma $\beta=100$. We instead fix its amplitude from an order-of-magnitude estimate with $\beta=100$ at the torus centre; the resulting field is weaker, with a volume-averaged value over the field-bearing region ($5.1\lesssim r\lesssim7.3$, $|\theta-\pi/2|\lesssim0.18$) of
\begin{equation}\label{eq:beta}
  \beta_0 \equiv \frac{\int_V P_{\rm eff}~{\rm d}V}{\int_V (B^2/2)~{\rm d}V} \approx 202 .
\end{equation}

The domain spans $2\le r\le275.6$ (log-spaced), $0\le\theta\le\pi$ and $0\le\phi<2\pi$, on a $64^3$ root grid with static mesh refinement (L$n$: $n$ levels beyond root). The torus body ($4\lesssim r\lesssim9.3$, $|\theta-\pi/2|\le2h_0$) is at L4, because the MRI quality factor $Q_\theta$ is only marginal at L3 \citep{Sano2004,Hawley2011}, and the cavity ($2.33\lesssim r<4$) is at L3 (Appendix~\ref{A:mri} assesses the MRI resolution in both); Fig.~\ref{fig:grid} shows the cells per scale height $H=h_0r$ at every radius. We call this setup L3C. Appendix~\ref{A:L2C} tests the sensitivity to cavity resolution by repeating the MHD run with a coarser cavity (L2C) and restarting the HD run with a finer cavity. The radial boundaries are one-way outflow, with the magnetic field zero-gradient; the $\theta$ boundaries are polar and $\phi$ is periodic.

We compute the gravitational torque of the gas on each MBH,
\begin{equation}
   T_{i}(t) = \hat{\mathbf z}\cdot\Bigl[\mathbf r_{\rm BH,i}\times \mathbf F_i\Bigr],\qquad
   \mathbf F_i = -\int_V \rho~\mathbf g_i~{\rm d}V,
\end{equation}
and $T_{\rm grav}=T_1+T_2$, integrated over the domain $r\ge2$. We also use the component kinetic and magnetic energies $\mathrm{KE}_i=\int\tfrac12\rho v_i^2~{\rm d}V$ and $\mathrm{ME}_i=\int\tfrac12B_i^2~{\rm d}V$; we use $\mathrm{ME}_r/\mathrm{ME}_\phi$ as a proxy for MRI saturation (Appendix~\ref{A:mri}).

We save 3D snapshots every $2.5~\tbin$, an integer number of binary half-periods, so every snapshot samples the same binary phase; time-resolved quantities are therefore taken from the history output. Restarts of both runs with snapshots every $2.25~\tbin$ over $\gtrsim100~\tbin$, which step the binary phase by $\pi/2$ (Appendix~\ref{A:mdot}), give the same time-averaged surface density, cavity eccentricity and $\alpha_{\rm eff}$ to within a few percent, so the snapshot averages are not phase-biased. The history output records $M_{\rm gas}$, the mass flux through $r=2.5$, $T_i$, $\mathrm{KE}_i$ and $\mathrm{ME}_i$ every $0.1~\tbin$.

\section{Post-Processing Analysis}\label{S:Post}
\begin{figure}
    \centering
    \includegraphics[width=\linewidth]{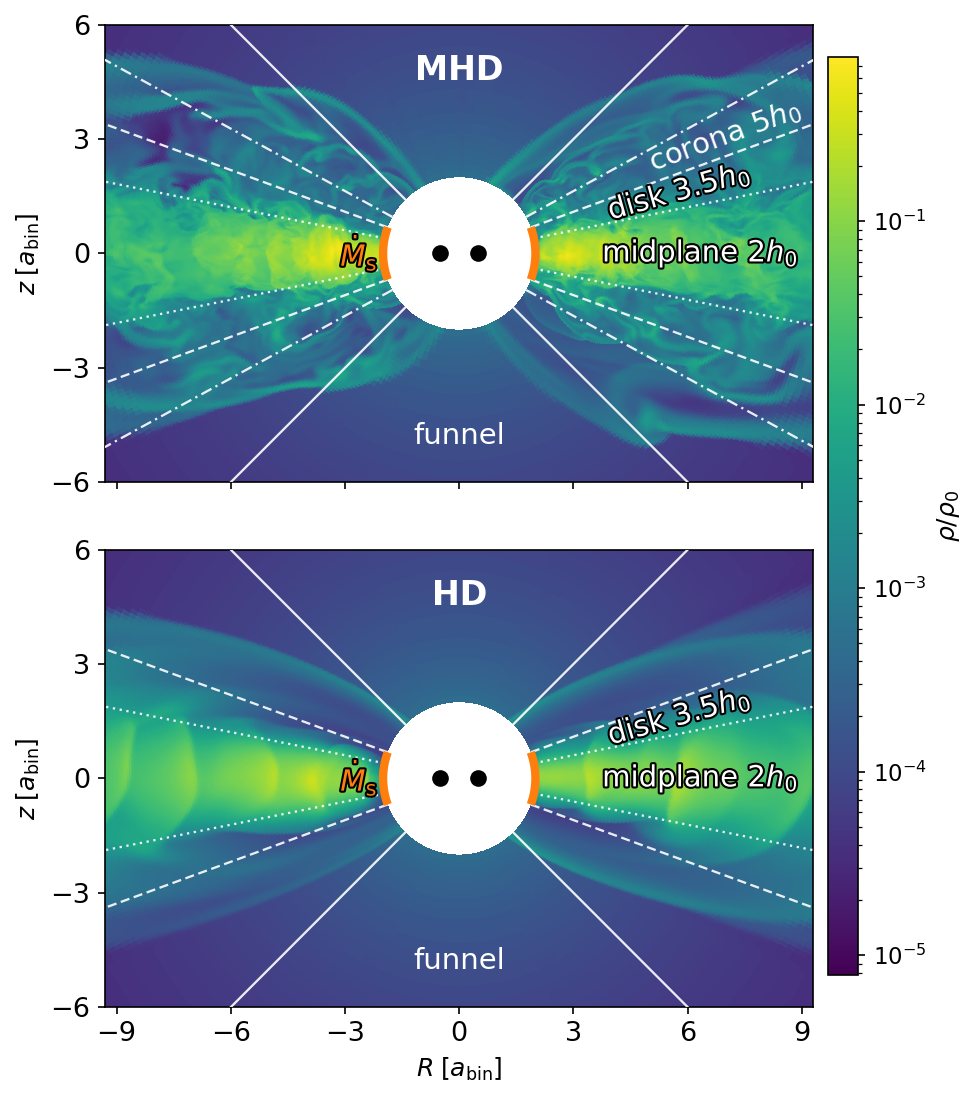}
    \caption{Edge-on density slices of the MHD (top) and HD (bottom) runs at $t=650~\tbin$, through $\phi=0$ (right half) and $\phi=\pi$ (left half), in cylindrical coordinates $R=r\sin\theta$, $z=r\cos\theta$, on a logarithmic scale over five decades below the initial torus peak density $\rho_0$. Lines mark the latitudes $|\theta-\pi/2|$ used by the diagnostics: the midplane window at $2h_0$ for $\alpha_{\rm eff}$ (dotted), the disk surface at $3.5h_0$ (dashed), the top of the MHD corona at $5h_0$ (dash-dotted) and the funnel boundary at $45^\circ$ (solid). The orange arcs on the inner boundary $r=2$, inside the disk surface, mark where the stream flux $\dot M_{\rm s}$ is measured. The white disk is the excised region and the dots mark the MBHs.}
    \label{fig:edgeon}
\end{figure}

The total gas mass changes through the mass flux across the inner boundary, $\dot M_{\rm in}$, and through mass injected by the floor and ceiling, $\dot M_{\rm inj}$: $\dot M_{\rm in}=-dM_{\rm gas}/dt+\dot M_{\rm inj}$. Floor resupply of the outer atmosphere ($r>50$) supplies $\approx65\%$ of the mass crossing the inner boundary in both runs; it reaches the inner boundary as a steady, high-latitude infall that carries negligible torque (Appendix~\ref{A:mdot}). We therefore define the accretion rate of torus gas as $\dot M\equiv-dM_{\rm gas}/dt$. When normalizing the torque we use instead the stream flux $\dot M_{\rm s}$ through the inner boundary within $|\theta-\pi/2|<3.5h_0$, measured at history cadence to be $\dot M_{\rm s}=1.22~\dot M$ (MHD) and $1.6~\dot M$ (HD) (Appendix~\ref{A:mdot}). Figure~\ref{fig:edgeon} marks these regions on the simulated density; Appendix~\ref{A:mdot} describes the mass flows between them. Three latitudes recur: $2h_0$, the extent of the finest refinement tier and of the window in which $\alpha_{\rm eff}$ is measured; $3.5h_0$, the minimum of the latitudinal mass flux, which separates the streams from the infall and defines $\dot M_{\rm s}$; and $5h_0$, the top of the magnetically supported corona that forms only in MHD (Appendix~\ref{A:mdot}).

We quantify the torque exerted by the gas on the binary through the dimensionless ratio
\begin{equation}\label{eq:xi}
  \xi\equiv \frac{T_{\rm grav}}{\dot M_{\rm s}~a_{\rm bin}^2\Omega_{\rm bin}},
\end{equation}
i.e., the gravitational torque per unit mass delivered to the binary through the streams, in units of the specific angular momentum of the binary orbit. This is
analogous to the angular-momentum eigenvalue commonly used in CBD studies \citep[e.g.,][]{Miranda2017,Duffell2024}, but
includes only the gravitational torque from the gas at $r\geq2~a_{\rm bin}$. We define the onset of a quasi-steady state,
$t_{\rm s,\xi}$, as the time after which the $100$-orbit moving average of $\xi$ no longer shows a secular trend, and we quote $\xi$ averaged from $t_{\rm s,\xi}$ to $1000~\tbin$. Here $\xi(t)$ is the ratio of the $100$-orbit means of $T_{\rm grav}$ and $\dot M_{\rm s}$, the latter from the mass lost over the window, not the mean of the instantaneous ratio, which the noisy $\dot M$ at $0.1~\tbin$ cadence would bias; $\langle\xi\rangle$ is the mean of this running series, whose first point falls at $t_{\rm s,\xi}+100~\tbin$.
Since our finite disk drains on the viscous time of the cavity edge, which exceeds the duration of our simulations, this state is quasi-steady rather than truly steady: $\dot M$ and $T_{\rm grav}$ may both still evolve while their ratio remains approximately constant, which is why we track the ratio rather than either quantity alone; a finite, draining disk approaches the eigenvalue of a steadily fed one only on that viscous time \citep{Munoz2020}.

To identify what drives angular momentum transport in each run, we decompose the $r\phi$-stress into its Reynolds, Maxwell, and viscous components \citep[e.g.,][]{Shi2012}:
\begin{equation}
  R_{r\phi}=\rho~\delta v_r~\delta v_\phi,~
  M_{r\phi}=-B_rB_\phi,~
  V_{r\phi}=-\Pi_{{\rm visc},r\phi},
\end{equation}
where $\delta v_i = v_i-\tilde v_i$ is the deviation from the mass-weighted azimuthal average $\tilde v_i=\langle\rho v_i\rangle_\phi/\langle\rho\rangle_\phi$ evaluated at each $(r,\theta)$, $\boldsymbol\Pi_{\rm visc}=\rho\nu\big[\nabla\mathbf v+(\nabla\mathbf v)^{T}-\tfrac23(\nabla\cdot\mathbf v)~\mathcal I\big]$ is the viscous stress tensor with $\nu(r)$ of Section~\ref{S:NumSetup}, $M_{r\phi}$ is nonzero only in the MHD run, and $V_{r\phi}$ only in the HD run. At each radius and snapshot, we form the effective viscosity parameter from the $\phi$- and polar-angle-window-integrated (denoted $\langle\cdot\rangle$) stress and pressure,
\begin{equation}\label{eq:alpha_eff}
  \alpha_{\rm eff}(r,t)=\frac{\langle R_{r\phi}\rangle+\langle M_{r\phi}\rangle+\langle V_{r\phi}\rangle}{\langle P_{\rm
eff}\rangle+\langle P_{\rm mag}\rangle},
\end{equation}
where $P_{\rm mag}=B^2/2$ (zero in the HD run). Including $P_{\rm mag}$ matters in the MHD run: the plasma beta (Eq.~\ref{eq:beta}), integrated over all $\theta$ in each radial band, saturates at averages of $3.4$ at the cavity edge and $2.2$ in the disk body. These column values are dominated by the magnetically supported layers above the disk; at the midplane $\beta\approx14$, and within the midplane window $P_{\rm mag}/P_{\rm gas}=0.2$ (Fig.~\ref{fig:pressure}). Each term in the numerator, divided by the same denominator, defines the corresponding contribution $\alpha_R(r,t)$, $\alpha_M(r,t)$, and $\alpha_V(r,t)$.

We report $\alpha_{\rm eff}$ and its components for two choices of polar-angle window. The \emph{midplane} window, $|\theta-\pi/2|\le2h_0$ (equivalently $|z|\lesssim2H$), isolates the local, near-midplane transport most directly comparable to standard shearing-box and thin-disk $\alpha$ measurements \citep[e.g.,][]{Hawley1995,Shi2012}. The \emph{column} window, $0\le\theta\le\pi$, integrates pole to pole and additionally captures the transport carried by the magnetized corona above the disk in the MHD run, a layer essentially absent in the field-free HD run. Following \citet{Tiwari2025a}, who compute $\alpha$ as a radial profile from shell-integrated stresses and pressures and only then report a single value averaged over the disk body, we average $\alpha_{\rm eff}(r,t)$ (and its $\alpha_R$, $\alpha_M$, $\alpha_V$ components) over the sampled snapshots at fixed radius,
then average over radius within each radial band ($r=2$-$4$ and $4$-$6~a_{\rm bin}$) and window.

Since the disk's self-gravity is neglected and the gas is isothermal, the simulations are scale-free: the disk mass, and hence the Eddington ratio, rescales $\dot M_{\rm s}$ and $T_{\rm grav}$ together and leaves $\xi$ and every other normalized quantity we report, such as $\alpha_{\rm eff}$ and $\Sigma/\Sigma_0$, unchanged.

\section{Results}\label{S:Result}

\begin{figure*}
    \centering
    \includegraphics[width=\textwidth]{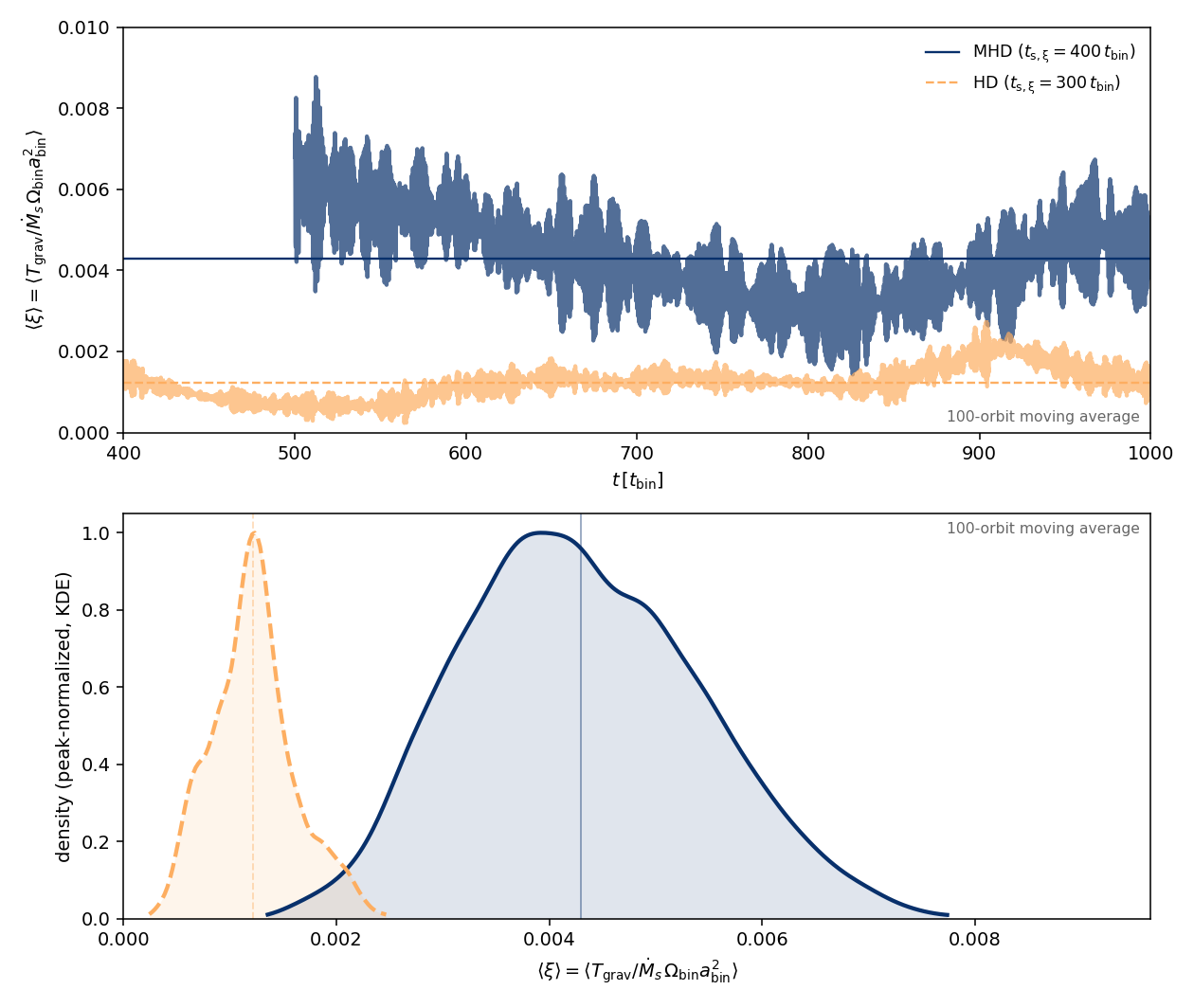}
    \caption{Torque normalization $\xi=T_{\rm grav}/(\dot M_{\rm s}~a_{\rm bin}^2\Omega_{\rm bin})$ (Eq.~\ref{eq:xi}) for the L3C MHD (solid) and HD (dashed) runs, with $\dot M_{\rm s}$ the stream flux through the inner boundary (Fig.~\ref{fig:edgeon}). Top: $100$-orbit moving average of $\xi$ from $t_{\rm s,\xi}=400~\tbin$ (MHD) and $300~\tbin$ (HD); the rapid variation is residual binary-phase torque passing through the boxcar window, and horizontal lines mark the time averages $\langle\xi\rangle=4.3\times10^{-3}$ (MHD) and $1.2\times10^{-3}$ (HD). Bottom: distribution of the moving-average values (Gaussian kernel density estimate, normalized to unit peak), with the time averages as vertical lines.}
    \label{fig:xi}
\end{figure*}

Figure~\ref{fig:xi} shows the torque normalization $\xi$ (Eq.~\ref{eq:xi}) in both runs. After $t_{\rm s,\xi}=400~\tbin$ (MHD) and $300~\tbin$ (HD), the $100$-orbit running mean oscillates without secular trend about its time average, $\langle\xi\rangle=4.3\times10^{-3}$ in MHD and $1.2\times10^{-3}$ in HD, a factor of $3.5$ apart. Each $\langle\xi\rangle$ is uncertain by about $10\%$ (MHD) and $20\%$ (HD), combining the statistical error of the time average with that of the stream-flux factor (Appendix~\ref{A:mdot}), so their ratio is $3.5\pm0.8$. The difference is one of torque, not of accretion rate: over the windows of $\xi$ the stream flux $\dot M_{\rm s}$ is the same in both runs to $2\%$ ($1.13\times10^{-10}$ against $1.15\times10^{-10}$), while $\langle T_{\rm grav}\rangle$ is $3.8$ times larger in MHD. The MHD run fluctuates more in absolute terms, but the relative fluctuations are similar, with standard deviations of $26\%$ (MHD) and $32\%$ (HD) of the mean. Table~\ref{tab:summary} collects the main numbers of the comparison.

\begin{table}
\centering
\caption{Summary of the L3C comparison. $\xi$ is averaged over $t_{\rm s,\xi}$-$1000~\tbin$, with $\sigma_\xi$ the standard deviation of its $100$-orbit running mean (Fig.~\ref{fig:xi}); all other quantities are averaged over $600$-$1000~\tbin$. $\alpha_{\rm eff}$ is the midplane-window value, the $\Sigma$ row is the MHD/HD ratio with its block-bootstrap error, and $e^{\rm eff}_{\rm cav}$ is the cavity-eccentricity proxy, the normalized $m=1$ Fourier amplitude of the annulus mass, with the standard deviation of its $100$-orbit running mean.}
\label{tab:summary}
\small
\setlength{\tabcolsep}{4pt}
\begin{tabular}{@{}lcc@{}}
\hline
 & MHD & HD \\
\hline
$\langle\xi\rangle$ & $4.3\times10^{-3}$ & $1.2\times10^{-3}$ \\
$\sigma_\xi$ & $1.1\times10^{-3}$ & $0.4\times10^{-3}$ \\
$\alpha_{\rm eff}$, cavity edge ($r=2$-$4$) & 0.05 & 0.17 \\
$\alpha_{\rm eff}$, disk body ($r=4$-$6$) & 0.11 & 0.14 \\
$\Sigma_{\rm MHD}/\Sigma_{\rm HD}$ ($r=2.5$-$3$) & \multicolumn{2}{c}{$3.4\pm0.1$ ($2$-$5$ at single epochs)} \\
$e^{\rm eff}_{\rm cav}$ ($r=2$-$4$) & $0.111\pm0.017$ & $0.064\pm0.009$ \\
$\delta\dot M/\dot M$ at $2\Omega_{\rm bin}$ & 0.06, intermittent & 0.2 \\
lump line / beat & yes / yes & transient / no \\
\hline
\end{tabular}
\end{table}

In both runs the torque exerted on the binary by the gas at $r\geq2~a_{\rm bin}$ is positive. This differs from HD CBD simulations that include the gas at $a_{\rm bin}\lesssim r<2~a_{\rm bin}$, in which the torque from the gas outside the binary orbit is negative \citep[see, e.g.,][]{Duffell2024}. We attribute the difference to the location of the principal outer Lindblad resonance of the binary potential \citep{Goldreich1979,Artymowicz1994}, the $m=2$ resonance at $r=(3/2)^{2/3}a_{\rm bin}\approx1.31~a_{\rm bin}$, which lies inside our excised region; in MHD simulations that include the gap, the torque density changes sign at $r\approx2~a_{\rm bin}$ \citep{Shi2012,Noble2012}.

As a check of the torque diagnostic against published results, we ran an HD variant with the inner boundary moved in to $r=\abin$ and $\alpha_{\rm vis}=0.1$, as in the Santa Barbara code comparison \citep{Duffell2024}. The torque from the gas at $r\ge\abin$, normalized by the accretion rate of torus gas $-dM_{\rm gas}/dt$, which excludes the floor-fed atmospheric infall (Appendix~\ref{A:mdot}), settles at $T_{\rm grav}=-0.64~\dot M\abin^2\Omega_{\rm bin}$ and persists at this value over $100$ orbits ($120$-$220~\tbin$), consistent with the $-0.62$ to $-0.70$ found by the three-dimensional codes in that comparison. The torque from $r\ge2\abin$ in our main runs excises the streams and minidisks and is not comparable to this benchmark.

\begin{figure*}
    \centering
    \includegraphics[width=\textwidth]{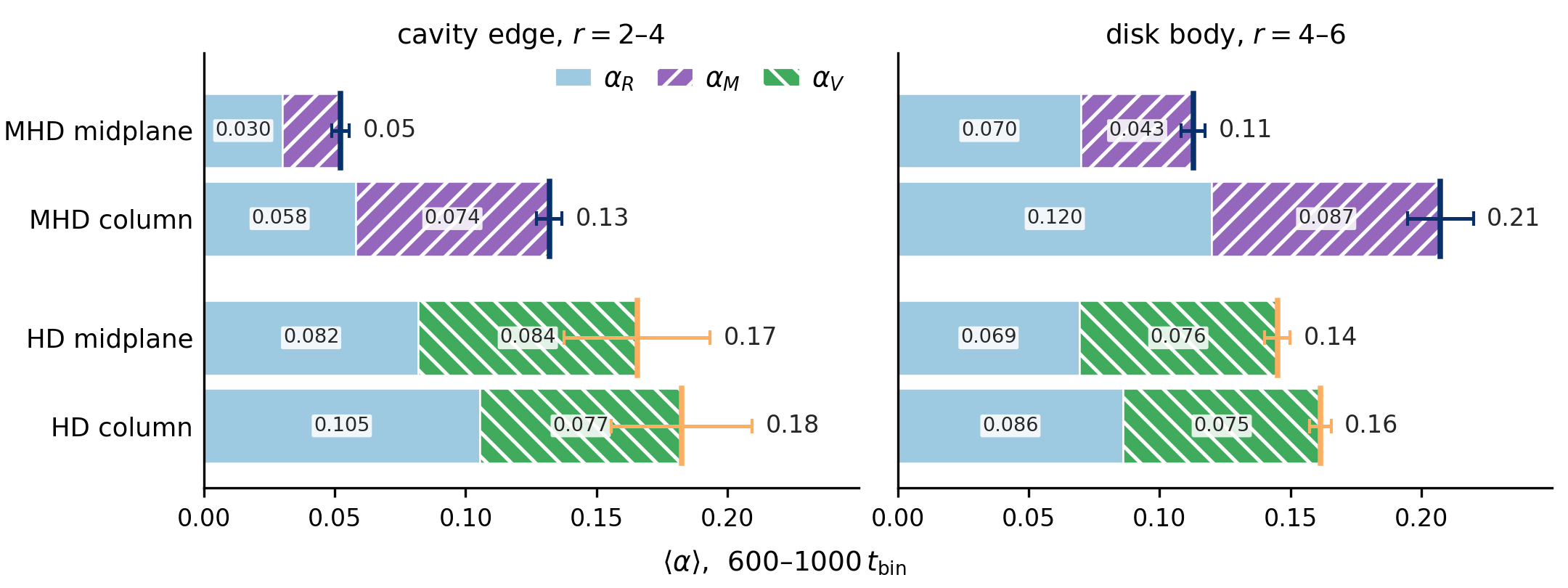}
    \caption{Time-averaged effective viscosity parameter $\langle\alpha_{\rm eff}\rangle$ (Eq.~\ref{eq:alpha_eff}) at the cavity edge ($r=2$-$4$, left) and in the disk body ($r=4$-$6$, right), over $600$-$1000~\tbin$ and over radius within each band. Each bar stacks the Reynolds ($\alpha_R$) and Maxwell ($\alpha_M$) contributions in MHD, and the Reynolds and viscous ($\alpha_V$) contributions in HD, with the total at its end. ``Midplane'' rows use the window $|\theta-\pi/2|\le2h_0$ (dotted line in Fig.~\ref{fig:edgeon}), ``column'' rows integrate over all $\theta$. Error bars are the standard deviation of the $100$-orbit running mean of the total, a measure of variability rather than of the error of the mean. The viscous contribution is $\alpha_V=\alpha_{\rm vis}~q~\bar\Omega/\Omega_K\approx1.5~\alpha_{\rm vis}$, with $q=-{\rm d}\ln\bar\Omega/{\rm d}\ln r$ the shear rate. The MRI is under-resolved at the cavity edge (Appendix~\ref{A:mri}), so the MHD values there are lower limits.}
    \label{fig:alpha}
\end{figure*}

Next, we compare the effective viscosity parameter $\alpha_{\rm eff}$ (Eq.~\ref{eq:alpha_eff}) of the two runs at the cavity edge ($r=2$-$4$, mostly at L3) and in the disk body ($r=4$-$6$, at L4), averaged over $600$-$1000~\tbin$ (Fig.~\ref{fig:alpha}). In the disk body the two runs transport angular momentum at similar rates: in the midplane window $\alpha_{\rm eff}=0.11$ (MHD) and $0.14$ (HD). At the cavity edge they differ strongly: $\alpha_{\rm eff}=0.05$ in MHD against $0.17$ in HD, a factor of $\approx3$, and the MHD value is a lower limit because the MRI is under-resolved there (Appendix~\ref{A:mri}). Over the full column, MHD transport roughly doubles in both bands ($0.13$ at the cavity edge and $0.21$ in the disk body), while HD changes little ($0.18$ and $0.16$): in MHD the magnetic pressure, nearly constant with height, takes over the vertical support at $|\theta-\pi/2|\approx1.8h_0$ and sustains a magnetically dominated ($\beta\lesssim0.1$) corona that holds $2.5$-$3.4$ times more gas than HD at $3.5$-$5h_0$, and this layer carries transport that has no HD counterpart (Appendix~\ref{A:mdot}, Fig.~\ref{fig:pressure}). This corona is not hot, since both runs are isothermal; it is gas held up by magnetic field that the MRI amplifies in the disk and that rises out of it \citep{MillerStone2000}. Near the midplane both disks remain locally isothermal Gaussians, with the $1/e$ density height at $\sqrt2h_0$ in both runs to within $2\%$; only at the $1/10$ level is the MHD disk $6\%$ thinner, where magnetic support takes over. In MHD the Reynolds stress exceeds the Maxwell stress in the midplane window, but it is not purely turbulent: removing the $m\le4$ azimuthal modes of the velocity leaves two thirds of the MHD body value ($0.048$ of $0.070$) and half of the HD one ($0.033$ of $0.069$), and at the cavity edge the HD Reynolds stress is almost entirely coherent spiral and stream flow ($0.070$ of $0.082$), equal to the viscous stress there. The cavity-edge contrast in $\alpha_{\rm eff}$ is therefore mostly the HD run's coherent flux and explicit viscosity, not a difference in turbulence.

\begin{figure}
    \centering
    \includegraphics[width=\linewidth]{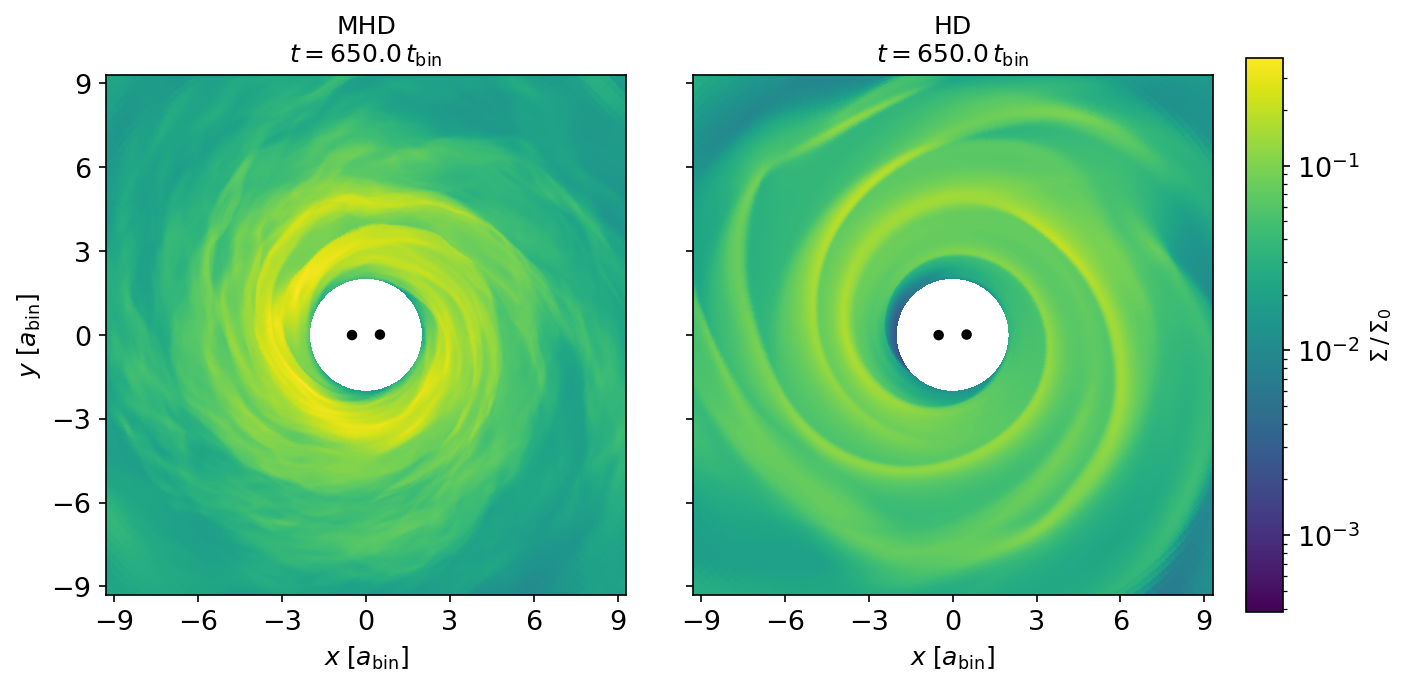}
    \caption{Surface density $\Sigma$ of the MHD (left) and HD (right) runs at $t=650~\tbin$, normalized by the maximum initial surface density $\Sigma_0=1.515\times10^{-8}$, on a shared logarithmic color scale.}
    \label{fig:disk}
\end{figure}

Figure~\ref{fig:disk} compares the surface density $\Sigma$ of the two runs at $t=650~\tbin$, within the window over which $\alpha_{\rm eff}$ is averaged. The MHD disk retains more mass near the inner boundary and shows more small-scale, filamentary structure, consistent with MRI-driven turbulence, whereas the HD disk exhibits pronounced spiral arms and a smoother radial mass distribution.

\begin{figure*}
    \centering
    \includegraphics[width=\textwidth]{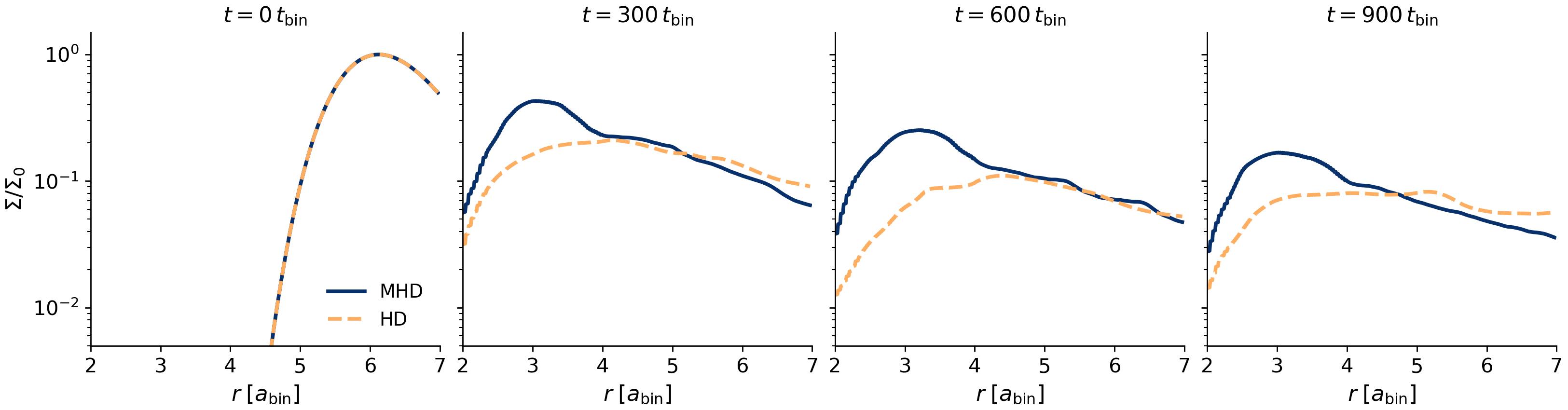}
    \caption{Azimuthally averaged surface density $\Sigma(r)$ of the MHD (solid) and HD (dashed) runs at $t=0$, $300$, $600$ and $900~\tbin$, normalized by $\Sigma_0$ as in Fig.~\ref{fig:disk}. At $t=0$ the torus profile falls below the plotted range inside $r\approx4.3$.}
    \label{fig:Sigma}
\end{figure*}

Figure~\ref{fig:Sigma} shows the azimuthally averaged surface density at $t=0$, $300$, $600$ and $900~\tbin$. Once the torus has spread, the two runs redistribute its mass differently, and the difference persists through the run. The MHD disk piles mass toward the cavity edge: at $r\approx2.5$-$3$ its surface density is $2$-$3$ times that of HD at $300$ and $900~\tbin$, and $4$-$5$ times at $600~\tbin$, when the inner HD disk is most depleted; averaged over $600$-$1000~\tbin$ the ratio is $3.4\pm0.1$. Beyond $r\approx5$ the two runs hold comparable mass, with the HD disk, which spreads its mass more evenly, $\approx15\%$ above MHD at $r=6$ at $300$ and $900~\tbin$. The pile-up is not an artifact of the excision radius (Appendix~\ref{A:boundary}).

Over $600$-$1000~\tbin$, the window of the quantities that follow, $\langle T_{\rm grav}\rangle$ is $3.7$ times larger in MHD. The torque from the gas at $r\ge2\abin$ is dominated by the cavity edge, where the gas is closest to the binary and non-axisymmetric, so it scales with the mass there and with its asymmetry. Both are larger in MHD. The annulus $r=2$-$4$ holds $2$-$3$ times more mass ($2$ at $300$ and $900~\tbin$, $3$ at $600~\tbin$; Fig.~\ref{fig:Sigma}). Its asymmetry, measured by the cavity-eccentricity proxy $e^{\rm eff}_{\rm cav}$, the normalized $m=1$ Fourier amplitude of the gas mass in the annulus \citep{Shi2012}, averages $0.111\pm0.017$ in MHD and $0.064\pm0.009$ in HD over $600$-$1000~\tbin$ (uncertainties are the standard deviations of the $100$-orbit running mean), $1.7$ times larger. The mass accounts for most of the contrast, since the torque scales with the annulus mass across both runs (predicted ratio $2.5$ against $3.7$); we attribute the remainder to the larger asymmetry. Both in turn follow from the weaker transport at the cavity edge in MHD, $\alpha_{\rm eff}=0.05$ against $0.17$ (Fig.~\ref{fig:alpha}), not from the residual difference in the disk body ($0.11$ against $0.14$): in HD simulations with $h_0=0.1$ the torque per unit accreted mass depends only weakly on the disk viscosity, while a lower viscosity at the cavity edge allows a larger, more eccentric cavity \citep{Dittmann2022}. Gas that reaches the cavity edge in MHD drains through it more slowly (inflow time $170~\tbin$ at $r=2.5$ against $40$ in HD), so it piles up there, and the cavity grows more eccentric; the same contrast between MRI-active and viscous disks is seen at the edges of planet-opened gaps \citep{Zhu2013}. The eccentricity contrast also depends on the inner boundary, since an HD test with the boundary at $r=\abin$ sustains an amplitude of $\approx0.2$ (Appendix~\ref{A:boundary}), so we do not interpret its magnitude as physical. Since the MRI is under-resolved at the cavity edge (Appendix~\ref{A:mri}), the MHD cavity-edge $\alpha_{\rm eff}$ is a lower limit, and a better-resolved cavity would reduce the pile-up; the L2C$\to$L3C comparison (Appendix~\ref{A:L2C}), in which the cavity-edge $\alpha_{\rm eff}$ rises by $50\%$ while $\xi$ agrees within $2\%$ on each run's own quasi-steady window, suggests the dependence is weak, but it is not established at convergence.

\begin{figure*}
    \centering
    \includegraphics[width=\textwidth]{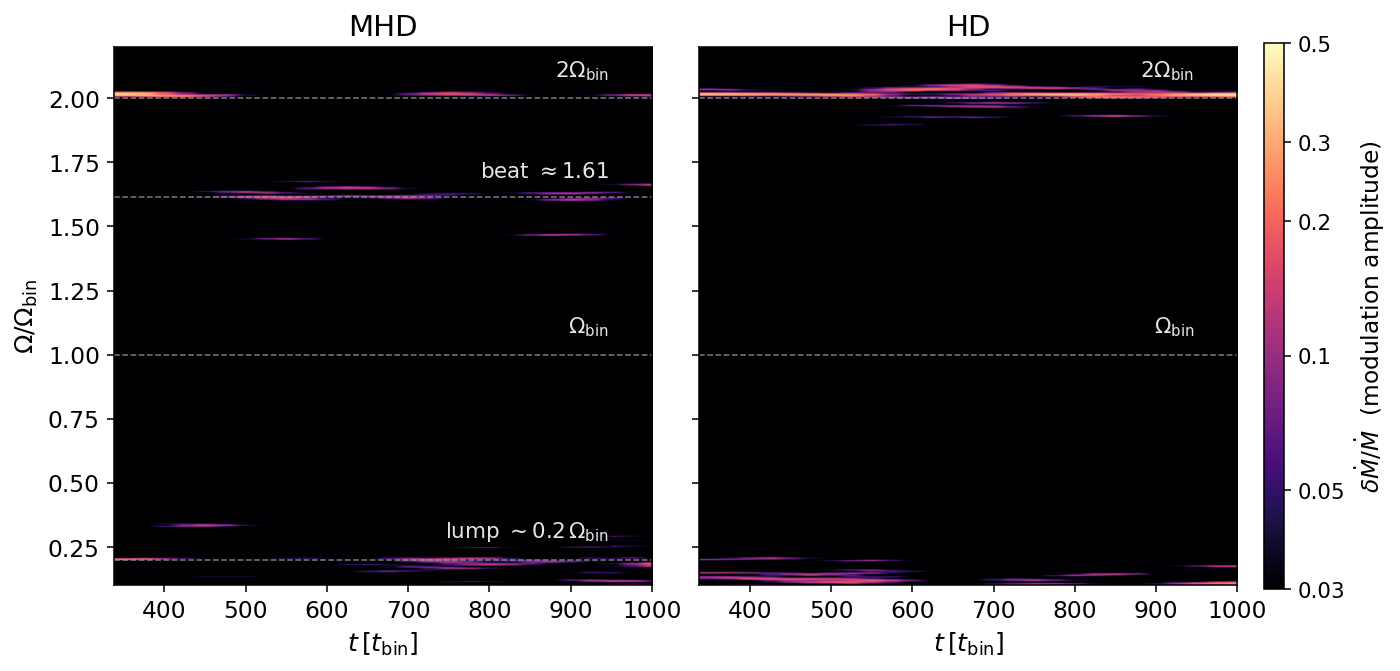}
    \caption{Lomb--Scargle spectrograms of the accretion rate $\dot M$ for the MHD (left) and HD (right) runs, from $100~\tbin$ windows stepped by $25~\tbin$ from $t=300~\tbin$. Color is the fractional modulation amplitude $\delta\dot M/\dot M$; only power above the $1\%$ false-alarm level in at least three consecutive windows is shown. Dashed lines mark the binary forcing frequency $2\Omega_{\rm bin}$, $\Omega_{\rm bin}$, the lump frequency $\Omega_{\rm lump}\approx0.2~\Omega_{\rm bin}$ and, for MHD, the beat $2\Omega_{\rm bin}-2\Omega_{\rm lump}\approx1.6~\Omega_{\rm bin}$.}
    \label{fig:mdot_ls}
\end{figure*}

Figure~\ref{fig:mdot_ls} shows the modulation of the accretion rate. The binary-forced line at $2\Omega_{\rm bin}$ is strong and persistent in HD ($\delta\dot M/\dot M\approx0.2$) but weak and intermittent in MHD ($\approx0.06$). The MHD run instead shows two lines that HD lacks: the lump orbital frequency $\Omega_{\rm lump}\approx0.2~\Omega_{\rm bin}$ (amplitude $\approx0.13$) and a beat at $2\Omega_{\rm bin}-2\Omega_{\rm lump}\approx1.6~\Omega_{\rm bin}$ (amplitude $\approx0.09$) that persists from $\approx400~\tbin$ to the end of the run, i.e.\ the two streams are modulated twice per lump orbit. In HD a lump appears only transiently, as a band at $0.05$-$0.2~\Omega_{\rm bin}$ over $\approx450$-$560$ and $\approx800$-$950~\tbin$ that tracks its $m=1$ cavity asymmetry, although lump-driven modulation at this frequency is a standard feature of HD simulations that resolve the cavity \citep{DOrazio2013,MunozLai2016}. The lump, a global density feature, also appears in the volume-integrated azimuthal kinetic energy of the MHD run throughout the run, at the $\lesssim2\%$ level (not shown). Runs with the inner boundary moved (Appendix~\ref{A:boundary}) show that the MHD lump and beat survive moving the boundary outward, whereas the absence of a persistent lump in HD is a consequence of the boundary at $r=2$ \citep[see, e.g.,][]{DOrazio2023}.

\section{Discussion}\label{S:Discussion}

What do these torques mean for the binary? For an equal-mass circular binary, the gravitational torque from the outer disk, $T_{\rm grav}=\xi~\dot M_{\rm s}\abin^2\Omega_{\rm bin}$, changes the separation by $\Delta a/a=8~\xi~\Delta M_{\rm s}/M_{\rm bin}$ \citep{Garg2024b}: $0.034\%$ (MHD) or $0.0096\%$ (HD) for every $1\%$ of the binary mass delivered through the streams. This is the outer-disk term alone. The full rate, $\dot a/a=2\dot J/J-3\dot M/M_{\rm bin}$ \citep{Duffell2024}, also contains the mass-gain term and the angular momentum of the accreted gas, both set inside the excised region, where the accretion torque and the $m=2$ outer Lindblad resonance make the net torque negative in simulations that resolve them \citep{Munoz2019,Duffell2024}, so the net sign of $\Delta a/a$ is not determined here. Our result is that the positive outer-disk term is $3.5$ times larger with MRI transport than with an $\alpha$-viscosity of the same disk-body $\alpha_{\rm eff}$, so the net torque in a full MHD calculation is expected to be less negative than $\alpha$-disk models predict, or to change sign at a different mass ratio or disk thickness. The absolute rates are set by the arbitrary disk mass: the torus (Appendix~\ref{A:mdot}) drains with an e-folding time of $1700~\tbin$ (MHD) and $2200~\tbin$ (HD), so the MHD disk accretes $27\%$ faster onto the inner boundary while delivering the same stream flux to the binary.

How does this compare with earlier MHD work? The only MHD simulation to report a torque per unit accreted mass is that of \citet{Shi2012}, whose domain extends in to $0.8\abin$ and so includes the gap region that dominates the torque: for the gas at $r\ge0.8\abin$ they find $T_{\rm grav}/(\dot M\abin^2\Omega_{\rm bin})\approx-0.67$ over $77$ orbits, within the range of the Santa Barbara comparison and close to the $-0.64$ of our HD variant with the boundary at $\abin$ (Section~\ref{S:Result}). This agreement should not be over-read: it compares an MHD disk cut at $0.8\abin$ and averaged over a few tens of orbits with an $\alpha=0.1$ HD disk cut at $\abin$, and the torque from the gap region is set by the resonant response at $r\approx1.3$-$1.6\abin$ rather than by the transport mechanism, so it checks the magnitude of the gap torque, not the equivalence of MHD and HD there. Neither they nor \citet{Noble2012} isolate this outer-disk contribution, and later MHD simulations \citep{Shi2015,Noble2021,Tiwari2025a} report accretion rates and their modulation rather than torques. The MHD/HD contrast in $\xi$ therefore has no direct counterpart in the literature. The closest statement is that of \citet{Shi2012}, who find a larger torque per unit disk mass in MHD than in an $\alpha$-disk because MHD stresses deliver more mass into the gap; here MHD transport places $2$-$3$ times more mass at the cavity edge. In both cases the magnetized disk puts more gas where the torque is exerted.

The difference in $\xi$ bears directly on the GW signal: the dephasing that the gas imprints on an inspiral in the LISA band scales with the gas torque, $\xi\dot M_{\rm s}$ \citep{Garg2024b}. Since the two runs deliver the same stream accretion rate $\dot M_{\rm s}$ (Section~\ref{S:Result}), the larger torque in MHD translates into a phase shift $3.5$ times larger than in the $\alpha$-disk model, if the contrast in the outer-disk torque carries over to the total gas torque, whose dominant terms arise inside $2\abin$. The variability of $\xi$ could matter as well: the distributions in Fig.~\ref{fig:xi} are close to Gaussian, with the MHD run three times broader in absolute terms. Because the dephasing is detectable only above a minimum phase shift of $8/{\rm SNR}$~rad \citep{Kocsis2011,Garg2022}, a broader distribution of torques at the same mean could place a larger fraction of systems above that minimum \citep{Garg2026a}.

Several limitations bound these conclusions. The excision leaves the sign of the net torque to a calculation with the binary inside the domain, and it also isolates the outer disk from the return of stream material: in simulations that resolve the gap, part of each stream is flung back to the cavity edge with added angular momentum \citep{Shi2012,Noble2012}, which drives the cavity eccentricity, and \citet{Shi2012} find that growth ceases once the boundary exceeds $1.3\abin$. Our boundary at $2\abin$ therefore measures the torque that the cavity-edge transport alone produces, without this feedback, in both runs equally. Since the feedback raises the eccentricity, and with it the torque, our $\xi$ values are conservative, and the contrast between the runs rests on the transport difference that the feedback does not touch; our HD test with the boundary at $\abin$, whose eccentricity reaches $0.2$ (Appendix~\ref{A:boundary}), shows how much the feedback can add. Whether it widens or narrows the MHD/HD contrast is the first question a run with the binary inside the domain should answer.

The binary is equal-mass, circular and fixed, the disk has $h_0=0.1$ and is locally isothermal, and the azimuthal pressure gradient retains the single sound speed, a larger effective aspect ratio for non-axisymmetric structure (Section~\ref{S:NumSetup}); since the torque depends on $h_0$ and mass ratio \citep{Tiede2020,Dittmann2022}, the $3.5$-fold contrast is specific to this configuration. Thermodynamics is the untested ingredient most likely to change it: the disk thickness, fixed at $h_0=0.1$ here, is in reality set by how the gas cools, and in an otherwise identical MHD disk radiation makes the cavity edge thinner, denser and less eccentric \citep{Tiwari2025a}: the first two raise the torque and the third lowers it, so which dominates can only be settled by long simulations with cooling; with physical cooling in GRMHD, the surface-density maximum at the cavity edge and the lump disappear within tens of orbits \citep{Kim2026}, whereas with target-entropy cooling the same run shows our lump and beat lines, and an $\alpha$-disk with the binary inside the domain sustains the lump for hundreds of orbits \citep{Garg2026b}. The lump is therefore robust to the transport mechanism and to the inner boundary in MHD, but not to the thermodynamics, and the variability signature we attribute to it is conditional on how the real disk cools. The seed field is a single poloidal loop without net vertical flux, so the magnetically arrested regime, in which the field itself can alter the torque \citep{Most2025,Wang2025}, is not sampled. Each run is a single realization of a finite, draining torus: the disk mass is arbitrary, the state is quasi-steady rather than steady, and $\langle\xi\rangle$ is uncertain by $10$-$20\%$ (Section~\ref{S:Result}). The numbers in this paper are therefore specific to one configuration, but the trends are not: MRI transport weakens at the cavity edge, mass piles up there, the cavity grows more eccentric, and the torque on the binary rises.

To return to the question posed in Section~\ref{S:Intro}, the following predictions of $\alpha$-disk simulations survive the switch to MRI-driven transport, for an equal-mass circular binary with $h_0=0.1$ and matched disk-body transport: the sign of the torque from the gas beyond $2\abin$ (positive), the near-midplane transport rate in the disk body ($\alpha_{\rm eff}\approx0.11$ against $0.14$), the rate at which gas is delivered to the binary through the streams, and the modulation of the accretion rate at the binary forcing frequency. What does not survive is the magnitude of the torque: the outer-disk torque per unit accreted mass is $3.5$ times larger in MHD, because MRI transport at the cavity edge is three times weaker than the $\alpha$-viscosity, which piles up $2$-$3$ times more mass at the cavity edge and makes the cavity more eccentric. Two further differences follow from the magnetized layers above the disk, which have no HD counterpart: a corona that deflects the atmospheric infall and doubles the column-integrated transport, and a persistent lump whose beat with the binary modulates the accretion rate at $2\Omega_{\rm bin}-2\Omega_{\rm lump}$. The MHD cavity-edge transport is a lower limit (Appendix~\ref{A:L2C}) and $\xi$ has not been tested against the inner boundary (Appendix~\ref{A:boundary}), so the magnitude of the MHD/HD contrast, not its sign, is what further work at higher cavity resolution and with a smaller excised region should test.

Compared with MRI turbulence, an $\alpha$-viscosity reproduces how fast a CBD feeds the binary, but not how hard it pushes or pulls. The next generation of simulations must therefore resolve the MRI, with the binary inside the domain and radiation setting the disk's thickness, to measure that push or pull and the variability that accompanies it, because the push or pull is what LISA will feel and the variability is what LSST and Roman will see.

\section*{Acknowledgments}
We acknowledge the initial help from Sean Ressler regarding running \texttt{Athena++}. We acknowledge useful discussions with Mark Avara, Yoonsoo Kim, Andrew MacFadyen, Lucio Mayer, and Lizhong Zhang. We further acknowledge the use of the NYU IT HPC Torch cluster. MG acknowledges support from the Swiss National Science Foundation (SNSF) Postdoc Mobility fellowship P500-2\_235363. The authors also acknowledge the use of Claude Sonnet 5 \citep{Sonnet2026} and Claude Opus 5.5 \citep{Opus2026} for assistance with code development, the analysis, and editing and revising the text; the authors reviewed all output and take full responsibility for the content. This work made use of the NumPy \citep{harris2020array}, SciPy \citep{SciPy2020} and Matplotlib \citep{Hunter2007} Python packages.

\bibliography{MHD}
\bibliographystyle{aasjournal}
\normalsize

\appendix
\twocolumngrid
\setcounter{figure}{0}                       
\renewcommand\thefigure{\thesection\arabic{figure}}

\section{Mass budget and the accretion rate}\label{A:mdot}
\setcounter{equation}{0}
\setcounter{figure}{0}

Of the initial gas mass $M_{\rm gas}\approx1.86\times10^{-6}$, the torus holds $9.9\times10^{-7}$ and the floor-density atmosphere the rest, almost all at $r>50$. The gas mass falls through the flux across the inner boundary, $\dot M_{\rm in}$, and rises through mass injected by the floor and, in MHD, the Alfv\'en-speed ceiling, $\dot M_{\rm inj}$:
\begin{equation}\label{eq:massbudget}
    \dot M_{\rm in} = -\frac{dM_{\rm gas}}{dt} + \dot M_{\rm inj}.
\end{equation}
The ceiling injects $\approx1.9\times10^{-9}$ through $1000~\tbin$, $0.1\%$ of the mass crossing the inner boundary. The floor is not negligible because the domain is large: the atmosphere at $\rho_{\rm floor}$ out to $r=275.6$ holds $0.87\times10^{-6}$, as much as the torus, and falls in steadily on its free-fall time ($\sim700~\tbin$ from $r\sim250$) while the floor refills it. Inside the disk, where the density is five decades higher, the floor never triggers. From the snapshot mass budget of $r>50$, whose mass changes by $\lesssim9\times10^{-8}$ while $1.05\times10^{-6}$ leaves it inward, the floor resupplies $\approx1.1\times10^{-6}$ over $1000~\tbin$ in both runs, $\approx65\%$ of the $\approx1.7\times10^{-6}$ crossing the inner boundary. The time-integrated history flux through $r=2.5$ agrees to $4\%$. While this atmosphere is in steady throughflow, the remainder is the accretion rate of torus gas, and we define $\dot M\equiv-dM_{\rm gas}/dt$.

The latitude profile of the inward mass flux (snapshots, $t>300~\tbin$) separates the two components. At $r=2.5$ the flux within $3.5h_0$ is the stream, $1.0\times10^{-10}$ in MHD, equal to $\dot M$, and the flux outside it is a steady infall of $1.85\times10^{-10}$ (MHD) and $1.6\times10^{-10}$ (HD), close to the floor resupply rate; the minimum between the two at $3.5h_0$ is what defines the disk surface. Farther out the two runs differ: at $r=6$ and $10$ the MHD infall peaks at $4$-$7h_0$, above the corona of density $\sim10^{-11}$ that overlies the disk at $3$-$5h_0$, and passes over it into the polar funnel, whereas in HD, where the density there is $4$-$11$ times lower, it peaks at $2$-$5h_0$, on the disk surface, and merges with the disk. This is why the MHD stream flux equals $\dot M$ while the HD stream flux is $1.6\dot M$, and why the infall carries negligible torque: it holds only $2$-$8\%$ of each shell's mass at $r\le6$, and its gravitational torque is $1$-$2\%$ of the instantaneous torque in both runs (see Fig.~\ref{fig:edgeon} for the regions).

To measure the stream flux at the inner boundary, we restarted both runs at $600~\tbin$ with history columns that split the Riemann mass flux through $r=2$ at $|\theta-\pi/2|=3.5h_0$, every $0.1~\tbin$, for $91~\tbin$ (MHD) and $157~\tbin$ (HD). The flux inside, $\dot M_{\rm s}$, tracks $\dot M$ (correlation $0.95$) at a nearly constant ratio $\dot M_{\rm s}/\dot M=1.22\pm0.06$ (MHD) and $1.6\pm0.2$ (HD); the flux outside is steady to $3$-$6\%$. In HD the excess is merged infall; in MHD it is partly infall that joins the disk surface at $6\lesssim r\lesssim10$ and partly gas entering the streams inside the cavity. In MHD the cell-centred snapshot flux $\rho v_r$ under-reads this boundary flux, because the wedge flow reverses sign from cell to cell at the cavity edge, so we rely on the Riemann flux. The MHD factor is uncertain by $\approx10\%$.

\begin{figure}
    \centering
    \includegraphics[width=\linewidth]{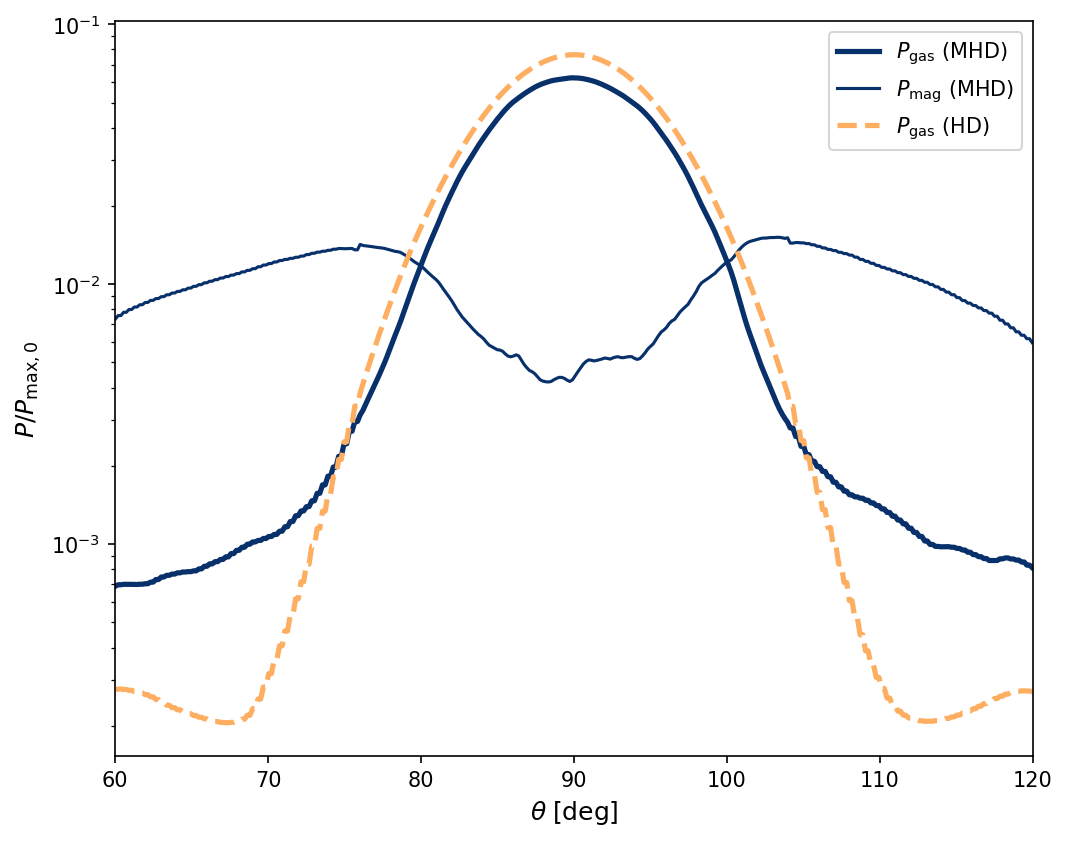}
    \caption{Vertical pressure structure at $r=6$: gas pressure in the MHD (thick solid) and HD (dashed) runs and magnetic pressure in MHD (thin solid) against polar angle, averaged over azimuth and over $602$-$650~\tbin$ (20 snapshots), normalized by the maximum gas pressure of the initial torus. The magnetic pressure is computed cell by cell before averaging, so the turbulent field contributes.}
    \label{fig:pressure}
\end{figure}

Figure~\ref{fig:pressure} shows the vertical structure behind this corona and behind the column-window excess of $\alpha_{\rm eff}$ in MHD (Section~\ref{S:Result}). At $r=6$, averaged over azimuth and over $602$-$650~\tbin$, the gas pressure of the two runs is similar at the midplane (HD exceeds MHD by $23\%$), where the magnetic pressure is only $7\%$ of the gas pressure; being nearly constant with height, the magnetic pressure exceeds the gas pressure tenfold at $3.5$-$5h_0$, where the HD gas pressure has fallen two decades below its midplane value. Transport in these magnetically supported layers is what the column window adds in MHD and not in HD.

\section{Sensitivity to cavity resolution}\label{A:L2C}
\setcounter{equation}{0}
\setcounter{figure}{0}
To assess the sensitivity of our results to the resolution of the cavity
region, we repeat the MHD simulation with a second grid setup, L2C, which is identical to L3C (Section~\ref{S:NumSetup}; Fig.~\ref{fig:grid}) except for the refinement level in the cavity. We vary only the cavity because it is where the MRI is under-resolved and where the MHD/HD contrast arises, whereas the disk body is at L4, where the MRI is resolved (Appendix~\ref{A:mri}); refining the whole torus by a further level would raise the cost by an order of magnitude. In L3C, the cavity $2.33\lesssim r<4$ is uniformly refined to L3; in L2C, this same radial range instead steps down through L1$\to$L2$\to$L3, transitioning at $r\approx2.7$ and $r\approx3.7$, so that only the intermediate band $2.7\lesssim r\lesssim3.7$ is uniformly one level coarser (L2) than L3C. Within this band, L2C is a factor of two coarser than L3C in each direction, resolving the local scale height with $\approx5$ radial, $8$ polar, and $4$ azimuthal cells. Inward of $r\approx2.7$, L2C is at L1, two levels coarser than L3C over $2.33\lesssim r\lesssim2.7$ and one level coarser at $2\le r\lesssim2.33$, where L3C is itself at L2. Outward of $r\approx3.7$, up to the L4 torus body at $r=4$, L2C already reaches L3 and is identical to L3C. The field-bearing region of the initial magnetic field lies entirely within the L4 tier, so the initial field configuration, including $\beta_0\approx202$ (Eq.~\ref{eq:beta}), is identical in the two setups. Through $1000~\tbin$, the cumulative ceiling-injected mass in the L2C MHD run is
$\approx0.09\%$ of the cumulative accreted mass. We do not use an L2C HD run: with only $2$-$4$ azimuthal cells per $H$ inside $r\approx3.7$, the L2C cavity under-resolves the HD cavity flow. Its torque nevertheless agrees with L3C: $\xi=1.18\times10^{-3}$ against $1.28\times10^{-3}$ over $400$-$1000~\tbin$, with the same stream-flux factor ($1.57$), so together with the refined-cavity restart below the HD torque is bracketed within $10\%$ from both sides, as expected when an explicit viscosity sets the transport. We test the HD run instead with a restart of the L3C HD run at $t\approx500~\tbin$ that raises the cavity $2.33\lesssim r<4$ to L4 and runs to $791~\tbin$ (Fig.~\ref{fig:grid}). This run was set up with adaptive mesh refinement (AMR), but its refinement criterion never triggered, so its mesh is static throughout; we refer to it as the refined-cavity restart.

\begin{figure*}
    \centering
    \includegraphics[width=\textwidth]{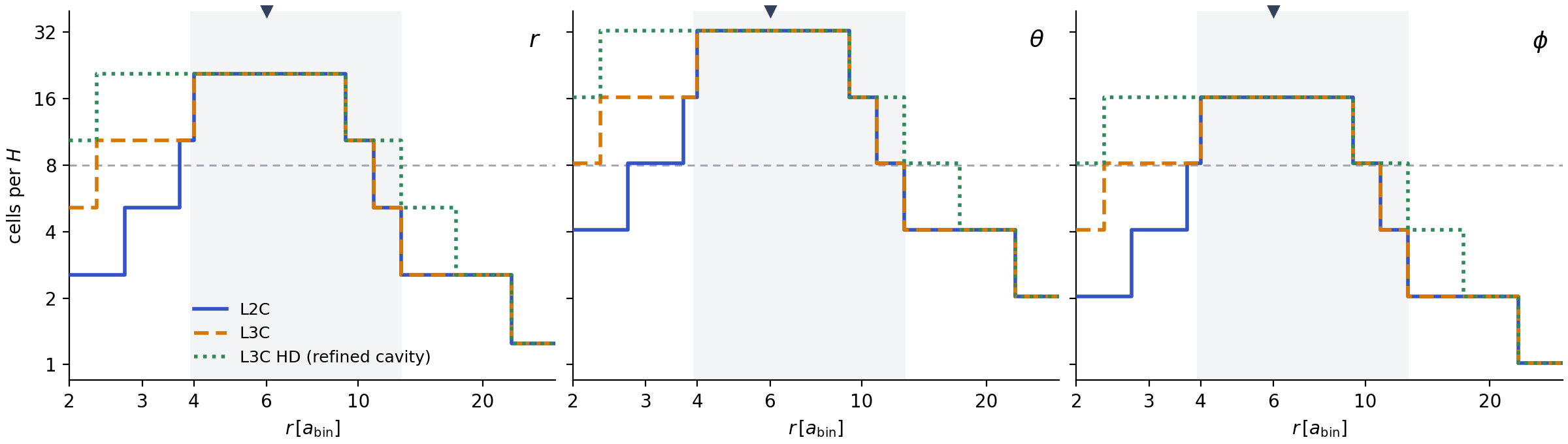}
    \caption{Midplane resolution of the L2C (solid) and L3C (dashed) setups and of the refined-cavity restart of the L3C HD run (dotted) in the radial (left), polar (middle) and azimuthal (right) directions, in cells per local scale height $H=h_0r$, from the meshblock refinement levels (at $t=0$ for L2C and L3C; the restart mesh is static). L2C and L3C differ only at $r<4$; the restart is finer than L3C in the cavity and at $9.3\lesssim r\lesssim17$. All three are at root resolution from $r\approx23.5$ to the outer boundary at $r=275.6$, beyond the plotted range. The shaded band marks the midplane extent of the initial torus and the triangle its pressure maximum at $r=6$; the horizontal dashed line marks 8 cells per $H$, roughly the minimum to resolve the MRI.}
    \label{fig:grid}
\end{figure*}

\begin{figure}
    \centering
    \includegraphics[width=\linewidth]{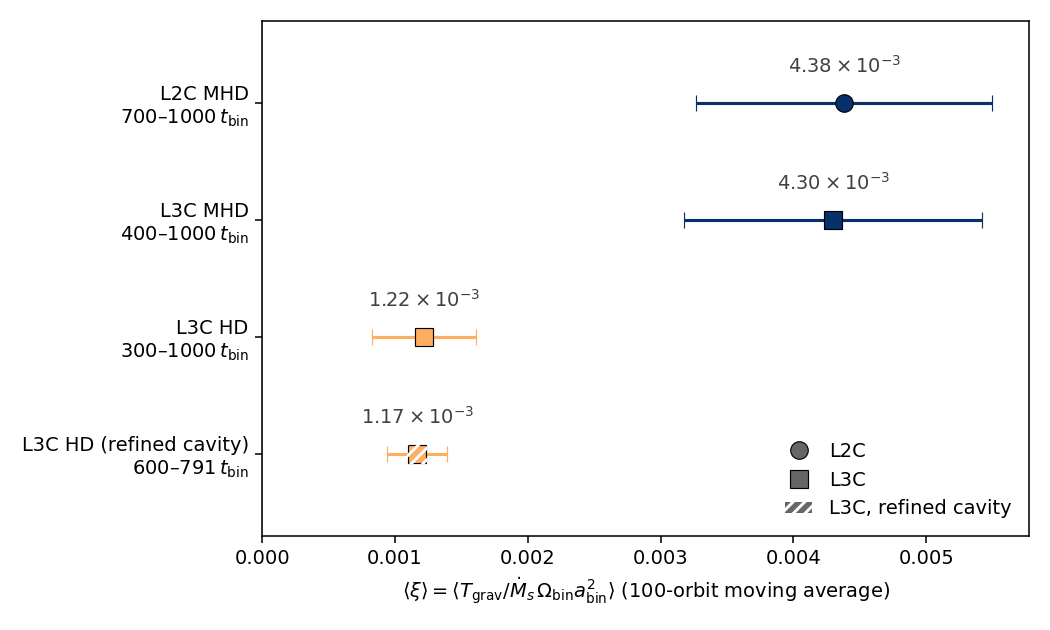}
    \caption{Time-averaged torque normalization $\langle\xi\rangle$ for L2C MHD (circle), the L3C runs (squares) and the refined-cavity restart of the L3C HD run (hatched), each over the window given below its label. Error bars as in Fig.~\ref{fig:alpha}.}
    \label{fig:xi_res}
\end{figure}

Figure~\ref{fig:xi_res} compares the torque normalization $\langle\xi\rangle$ (Eq.~\ref{eq:xi}) between the setups, each averaged over its own quasi-steady window. L3C MHD saturates earlier ($t_{\rm s,\xi}=400~\tbin$) than L2C MHD ($700~\tbin$). L2C MHD gives $\langle\xi\rangle=4.38\times10^{-3}$ against $4.30\times10^{-3}$ in L3C, a difference of $2\%$, far smaller than the $25$-$26\%$ spread of the $100$-orbit moving average in each run. The L2C MHD value uses its own stream-flux factor, $\dot M_{\rm s}/\dot M=1.30$, measured as for L3C from a $30~\tbin$ restart at $800~\tbin$ with the boundary-flux history columns; it exceeds the L3C factor of $1.22$ because the coarser cavity lets more off-stream gas enter the wedge inside $r=2.5$. For HD, the refined-cavity restart gives $1.17\times10^{-3}$ over $600$-$791~\tbin$, against $1.22\times10^{-3}$ for L3C HD. The MHD/HD contrast in $\xi$ is therefore insensitive to the cavity resolution.

\begin{figure*}
    \centering
    \includegraphics[width=\textwidth]{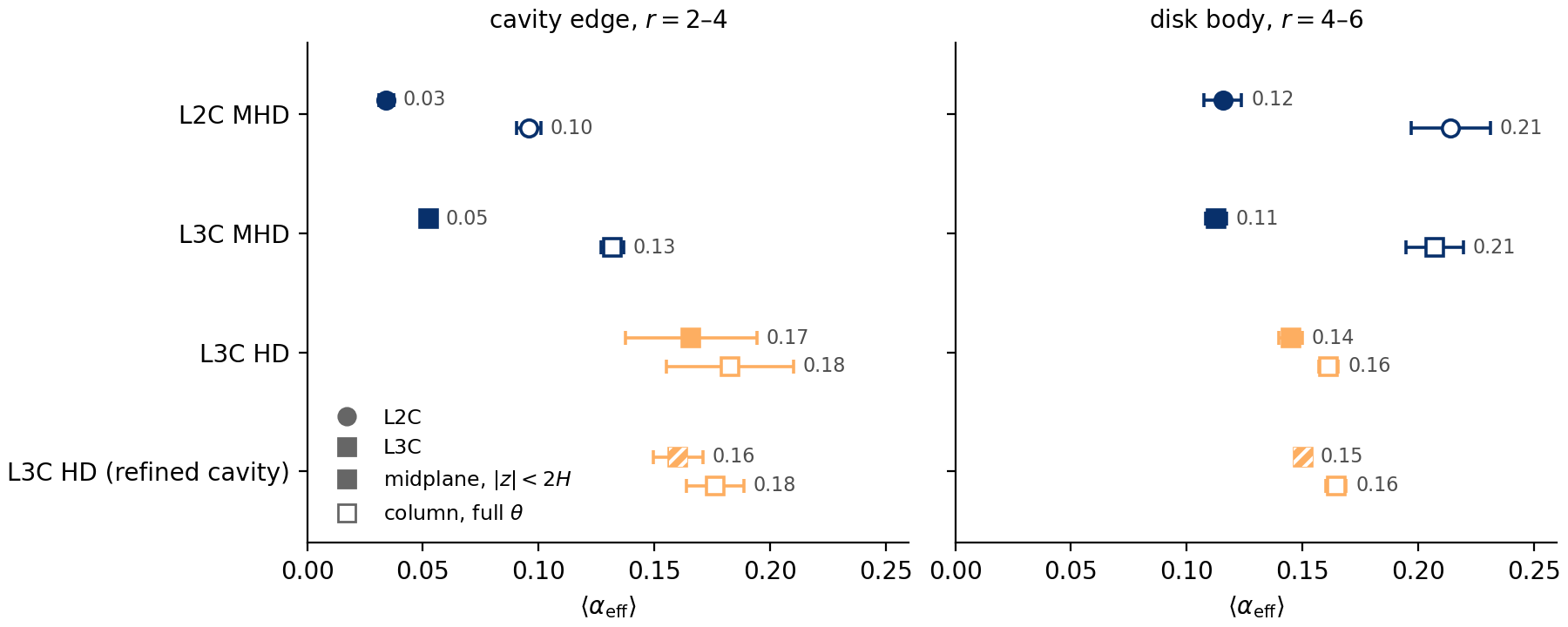}
    \caption{Time-averaged effective viscosity parameter $\langle\alpha_{\rm eff}\rangle$ (Eq.~\ref{eq:alpha_eff}) at the cavity edge ($r=2$-$4$, left) and in the disk body ($r=4$-$6$, right) for L2C MHD (circles), the L3C runs (squares) and the refined-cavity restart of the L3C HD run (hatched squares), over radius within each band and over $600$-$1000~\tbin$ ($600$-$790~\tbin$ for the restart). Filled markers use the midplane window and open markers the full column, as in Fig.~\ref{fig:alpha}; the numbers give each mean and error bars are as in Fig.~\ref{fig:alpha}. The MRI is under-resolved at the cavity edge in both MHD setups (Appendix~\ref{A:mri}).}
    \label{fig:alpha_res}
\end{figure*}

Figure~\ref{fig:alpha_res} makes the same comparison for $\langle\alpha_{\rm eff}\rangle$ (Eq.~\ref{eq:alpha_eff}) over $600$-$1000~\tbin$ ($600$-$790~\tbin$ for the refined-cavity restart), in the disk body ($r=4$-$6$) and at the cavity edge ($r=2$-$4$), for the body and column windows. In the disk body, which is at L4 in both setups, L2C and L3C MHD agree to $4\%$ in both windows ($0.116$ and $0.214$ against $0.113$ and $0.207$, midplane and column window respectively). At the cavity edge, where the setups differ, L2C MHD gives $\alpha_{\rm eff}=0.034$ (midplane) and $0.096$ (column), $\approx30\%$ below L3C ($0.052$ and $0.132$), several times the spread of each run: the coarser cavity carries less MRI stress. In HD the refined-cavity restart and L3C agree to within $4\%$ in both bands and both windows: $0.160$ and $0.176$ against $0.166$ and $0.183$ at the cavity edge, and $0.150$ and $0.165$ against $0.145$ and $0.161$ in the disk body, as expected for transport set mainly by the explicit viscosity. The midplane-window values used in the main text are thus converged with respect to the cavity resolution, while the MHD cavity-edge stress is a lower limit, at L2C and also at L3C, where the cavity under-resolves the MRI (Appendix~\ref{A:mri}).

\section{Sensitivity to the inner boundary}\label{A:boundary}
\setcounter{equation}{0}
\setcounter{figure}{0}
Two variants of the L3C runs with a different inner boundary test which results depend on the excision radius $r=2$. In an MHD run with the boundary moved out to $r=2.5$, otherwise identical and evolved to $527~\tbin$, the surface-density peak moves outward to $r\approx3.5$, following the new boundary, but at $t=300~\tbin$ the surface density at $r=3$-$3.5$ still exceeds that of HD by $20$-$40\%$, and the mass within $r=2.5$-$7$ is within $5\%$ of that with the boundary at $r=2$: the pile-up at the cavity edge (Fig.~\ref{fig:Sigma}) is not an artifact of the excision radius. In the spectrogram of its accretion rate (not shown) the lump line and its beat persist throughout, the beat at $\approx1.68~\Omega_{\rm bin}$ as $2\Omega_{\rm bin}-2\Omega_{\rm lump}$ requires for $\Omega_{\rm lump}\approx0.16~\Omega_{\rm bin}$, while the binary line at $2\Omega_{\rm bin}$ is absent: with the boundary at $2.5$ the streams no longer reach the domain edge in a binary-locked pattern.

The HD variant with the boundary moved in to $r=\abin$ is the benchmark run of Section~\ref{S:Result} ($\alpha_{\rm vis}=0.1$, evolved to $434~\tbin$). Its binary line is present throughout, and a lump band at $0.2$-$0.25~\Omega_{\rm bin}$ is the strongest feature of any run, with $\delta\dot M/\dot M$ approaching unity, accompanied by combination lines at $2\Omega_{\rm lump}\approx0.42~\Omega_{\rm bin}$ and at $2\Omega_{\rm bin}-\Omega_{\rm lump}\approx1.78~\Omega_{\rm bin}$ rather than the $2\Omega_{\rm bin}-2\Omega_{\rm lump}$ beat of MHD. Its cavity-eccentricity proxy sustains $\approx0.2$ at $r=2.5$-$4$, above both L3C runs. The absence of a persistent lump in our HD run is therefore a consequence of the boundary at $r=2$, whereas the MHD lump and beat survive moving the boundary outward. Because this variant differs from the HD run in viscosity and duration as well as in the boundary, it constrains the lump and the eccentricity, not $\xi$. Two external results point the same way: in a 3D $\alpha$-disk with the binary inside the domain and the same thermodynamics, the lump persists for hundreds of orbits with $\delta\dot M/\dot M\approx0.25$ \citep{Garg2026b}, so an $\alpha$-viscosity does not by itself suppress it, and in GRMHD the lump survives target-entropy cooling but not physical cooling \citep{Kim2026}, the thermodynamic counterpart of these boundary tests.

\section{MRI resolution}\label{A:mri}
\setcounter{equation}{0}
\setcounter{figure}{0}
We measure how well the MHD run resolves the MRI with the quality factors
\begin{equation}\label{eq:Q}
  Q_i=\frac{\lambda_i}{\Delta x_i},\qquad \lambda_i=2\pi\sqrt{\tfrac{16}{15}}~\frac{|v_{A,i}|}{\Omega},
\end{equation}
the number of cells across the fastest-growing MRI wavelength, where $v_{A,i}=B_i/\sqrt\rho$, $\Omega=(GM_{\rm bin}/r^3)^{1/2}$, and $\Delta x_\theta=r~\Delta\theta$ and $\Delta x_\phi=r\sin\theta~\Delta\phi$ are the true cell sizes at each cell's refinement level, following the definition of \citet{Tiwari2025a}. We use the polar and azimuthal factors, for which \citet{Hawley2011} recommend $Q_\theta\gtrsim10$ and $Q_\phi\gtrsim20$ (\citealt{Tiwari2025a} adopt $Q_\phi\gtrsim25$); we treat $Q_\theta=6$ \citep{Sano2004} and $Q_\phi=10$ as marginal. We complement them with three saturation diagnostics that approach resolution-independent values once the MRI is resolved \citep{Hawley2011}: the magnetic stress ratio $\alpha_{\rm mag}=\langle-B_rB_\phi\rangle/\langle B^2/2\rangle\approx0.3$-$0.4$ (the Maxwell stress of Eq.~\ref{eq:alpha_eff} normalized by the magnetic rather than the total pressure, so that $\alpha_M\approx\alpha_{\rm mag}/\beta$ at large $\beta$), the field-component ratio $\langle B_r^2\rangle/\langle B_\phi^2\rangle\gtrsim0.2$, and the field tilt angle $\tfrac12\arcsin\alpha_{\rm mag}\approx11$-$13^\circ$.

\begin{figure}
    \centering
    \includegraphics[width=\linewidth]{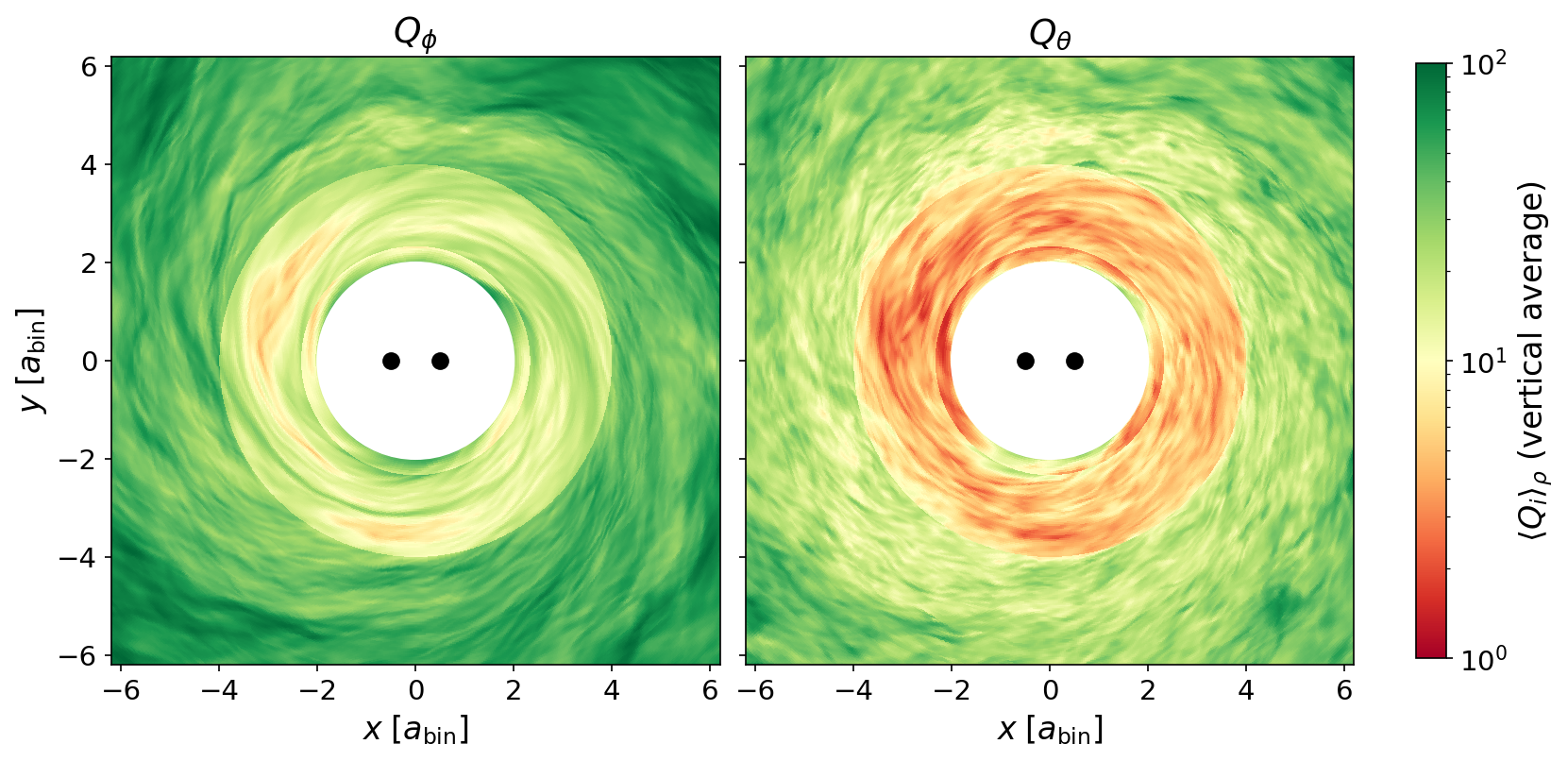}
    \caption{Face-on maps of the MRI quality factors $Q_\phi$ (left) and $Q_\theta$ (right; Eq.~\ref{eq:Q}) in L3C MHD at $t=650~\tbin$, density-weighted over $|\theta-\pi/2|\le\pi/4$, within $|X|,|Y|\le6.2~a_{\rm bin}$. The logarithmic color scale runs from red (under-resolved) through yellow ($Q\approx10$) to green (well resolved). The sharp edges at $r=2.33$ and $4$ are the L2/L3 and L3/L4 refinement boundaries; the white disk is the excised region and the dots mark the MBHs.}
    \label{fig:Q_faceon}
\end{figure}

Figure~\ref{fig:Q_faceon} shows face-on maps of $Q_\phi$ and $Q_\theta$ in L3C MHD at $t=650~\tbin$, in the format of \citet[their Fig.~18]{Tiwari2025a}. They show an under-resolved ring at $2\lesssim r\lesssim4$, where the refinement steps down from L4 to L3, surrounded by a well-resolved torus, whose density-weighted means over $r=4$-$8$ are $\langle Q_\phi\rangle=41$ and $\langle Q_\theta\rangle=21$ at this time. To quantify this we average $Q_\theta$ and $Q_\phi$ over azimuth and over $650$-$700~\tbin$ (21 snapshots), within the midplane window $|\theta-\pi/2|\le2h_0$ used for $\alpha_{\rm eff}$. The torus body ($r=4$-$6$, L4) is well resolved: the density-weighted means are $\langle Q_\theta\rangle=15.5$ and $\langle Q_\phi\rangle=31$, no gas has $Q_\theta<6$ or $Q_\phi<10$, and only $18\%$ has $Q_\theta<10$. The cavity edge ($r=2$-$4$, L3, and L2 inside $r=2.33$) is not: $\langle Q_\theta\rangle=3.0$ and $\langle Q_\phi\rangle=10$, with $93\%$ of the gas below $Q_\theta=6$. Since the disk has spread inward by this time, the cavity edge holds $82\%$ of the gas within $r=2$-$6$ and $|\theta-\pi/2|\le2h_0$. Over $r=2$-$9$ in the same window the saturation diagnostics are $\alpha_{\rm mag}=0.31$ and $\langle B_r^2\rangle/\langle B_\phi^2\rangle=0.23$, at the lower edge of the converged ranges, while the tilt angle of $8.9^\circ$ falls short of them. Split by band over $600$-$1000~\tbin$, $\alpha_{\rm mag}$ is $0.23$ at $r=2$-$3$, $0.32$ at $3$-$4$ and $0.34$ at $4$-$6$, with $\langle B_r^2\rangle/\langle B_\phi^2\rangle=0.10$, $0.12$ and $0.24$; in L2C the same bands give $\alpha_{\rm mag}=0.14$, $0.25$ and $0.34$. The innermost band is numerically suppressed and improves with resolution, whereas at $3$-$4$ the stress ratio is close to its converged value while the field geometry is not. The midplane plasma $\beta$ is also highest there, $51$ against $20$ at $2$-$3$ and $19$ at $4$-$6$, so part of the weak cavity-edge transport is physical: little field reaches the cavity edge.

The MRI is therefore resolved in the torus body, where we measure $\alpha_{\rm eff}$ for the main text, but under-resolved at the cavity edge. The magnetic stress there is a lower limit, consistent with the $\approx30\%$ drop in cavity-edge $\alpha_{\rm eff}$ when the cavity is coarsened further to L2C (Appendix~\ref{A:L2C}).

\end{CJK*}
\end{document}